\documentclass[11pt,preprint]{aastex}

\begin{document}
\title{Solar Multiple  Eruptions From a Confined Magnetic Structure}
\author{Jeongwoo Lee\altaffilmark{1}, Chang Liu\altaffilmark{2}, Ju Jing\altaffilmark{2}, Jongchul Chae\altaffilmark{1} }
\affil{1. Department of Physics and Astronomy, Seoul National University, Seoul 08826, Korea}
\affil{2. Space Weather Research Laboratory, New Jersey Institute of Technology, Newark, NJ 07102, U.S.A}

\begin{abstract}
How eruption can recur from a confined magnetic structure is discussed based on the {\it Solar Dynamics Observatory} (SDO) observations of the NOAA active region 11444, which produced three eruptions within 1.5 hours on March 27, 2012. The active region had the positive polarity magnetic fields in the center surrounded by the negative polarity fields around. Since such a distribution of magnetic polarity tends to form a dome-like magnetic fan structure confined over the active region, the multiple eruptions was puzzling.  Our investigation reveals that this event exhibits several properties distinct from other eruptions associated with magnetic fan structures: (i) a long filament encircling the active region was present before the eruptions; (ii) expansion of the open-closed boundary of the field lines after each eruption suggestive of the growing fan-dome structure, and (iii) the ribbons inside the closed magnetic polarity inversion line evolving in response to the expanding open-closed boundary.
It thus appears that in spite of multiple eruptions the fan-dome structure remained undamaged, and the closing back field lines after each eruption rather reinforced the fan-dome structure. We argue that the multiple eruptions could occur in this active region in spite of its confined magnetic structure because the filament encircling the active region was adequate for slipping through the magnetic separatrix to minimize the damage to its overlying fan-dome structure. The result of this study provides a new insight into the productivity of eruptions from a confined magnetic structure.

\end{abstract}


\section{INTRODUCTION}

Understanding how solar eruption proceeds is an important step in achieving the ultimate goal of solar physics and space weather forecast. Although some observations suggested that the properties of the trigger tend to correlate with solar eruption productivity, which mechanism dominates solar eruptions is still debatable (Forbes 2000). The answer may lie in the specific magnetic field configuration favorable for hosting those eruptions, but the magnetic structures of eruption productive active regions (ARs) are highly complex, and it is difficult to determine the key elements of the magnetic field present at the onset of eruptions from the imaging data (Schrijver 2009, Kusano et al. 2012). In this regard, it may be particularly important to study the magnetic structures producing multiple eruptions because we can see how a system once erupts and becomes ready for the next eruption (Luoni et al. 2007, Chandra et al. 2011, Chen et al. 2015).

Recurrent flares can easily be understood in terms of magnetic reconnection that repeats without destroying the overall magnetic configuration. If magnetic flux emerging from below collides with the overlying magnetic field lines and only partly consumes the magnetic field, the structure may not necessarily be altered significantly to recover itself (Yan et al. 2012, Yang et al. 2014). A well-studied  recent example is NOAA AR 12192 which produced many strong flares from the bipolar magnetic structure (Thalmann et al. 2015).
Multiple flares from a sustaining magnetic structure are also called homologous flares (Gary \& Moore 2004, Goff et al. 2007). {\bf Another type of flares with opposite character to the confined flares is {\it eruptive flares}. They are accompanied by coronal mass ejections (CMEs), and hardly repeat themselves over a short time scale of a few hours.}

Recurrent {\it eruptions} from a {\it confined} magnetic structure can be puzzling, {\bf since `eruption' and `confined' may sound conflict to each other. By the eruption we refer to an event involved with magnetic field ejection, and by the confined magnetic structure, a region covered by a fan dome-like magnetic separatrix. This should not be confused with {\it confined eruption} that is a term reserved for failed CMEs (Ji et al. 2009). In a successful eruption, also called   {\it breakout eruption} (Antiochos 1998, Shen et al. 2012) the overlying  separatrix is destroyed by the ejecta, and the magnetic system can hardly restore the pre-eruption configuration to launch another eruption within a short time scale of a few hours. For this reason {\it recurrent eruptions} from a confined magnetic structure may be considered impossible.
The aforementioned bipolar structure in the standard model (Forbes 2000) may not be subject to this problem, because its separatrix lies between open and closed field lines so that  an ejecta may escape along the separatrix, and the magnetic null point can still lie above the looptop to initiate the next eruption. Even a more complex structure may avoid this problem if it involves open field lines surrounding the closed fields and the eruption proceeds from outside to inside so as to progressively open the field lines (e.g., Moore et al. 2010). Only in the case that the erupting magnetic rope is inside a closed separatrix, it can be challenging to explain how such a magnetic structure can produce multiple eruptions.}

In this letter we study a special event worthy of attention in this context, the three consecutive eruptions that occurred on 27 March 2012  from the NOAA  AR 11444. The AR had one polarity magnetic flux isolated in the middle, surrounded by the other polarity field. Such an AR tends to form a fan-spine structure with a magnetic null point above the dome-like fan (Parnell et al. 1996). The reconnection in such a magnetic field structure typically results in a circular ribbon flare along the footprint of the fan (e.g., Masson et al. 2009). We will, however, show that the present event have different properties from those of the circular ribbon flares, and suggest that those differences are the key to understanding the multiple eruptions from the confined magnetic structure.
We describe the data in \S2, and analysis of the coronal images in \S3 and \S4. Based on the result, we offer an interpretation of the multiple eruptions in \S5, and conclude in \S6.

\section{Data}

We use the (E)UV images obtained from Atmospheric Imaging Assembly (AIA; Lemen et al. 2012) together with magnetic field information from the Helioseismic and Magnetic Imager (HMI; Schou et al. 2012) on board the {\it Solar Dynamics Observatory} ({\it SDO}; Pesnell et al. 2012). The AIA images are obtained with a cadence of 12 s and {\bf pixel size of} 0.6 arcsec. Our target is the NOAA AR 11444 at its location N21 W17 on March 27, 2012. The preflare activity started around 02:30 UT. The three eruption occurred at around 02:55 UT, 04:14 UT and 04:24 UT, respectively. Two flares with GOES class C5.3 and and C1.7 occurred at $\sim$02:52 UT and at $\sim$04:25 UT, respectively.
We followed the normal procedure to download the data at Level-1.0 and used aia-prep.pro routine available in SSW packages to update it to the Level-1.5. We set the reference time at 02:00 UT and used rot-xy.pro to co-align the images from all of the AIA channels, rescale them to a common plate scale and derotated the images. We finally normalized the intensities to 1-sec exposure time for all the wavelengths.

\section{The Eruptions}

Figure 1 shows the AIA images at 171 \AA\ in the top panels and the 304 \AA\ images together with an HMI magnetogram in the bottom. Four time intervals are chosen to represent the pre-flare, the first to the third eruptions, respectively.  In the pre-eruption stage (the first column) the most characteristic feature is the long filament (marked by `FF') encircling almost half of the AR circumference. It may be not a single filament, but a collection of 2 or 3 filaments. All eruptive activities occur along these filaments around the AR in the clockwise direction. In the first eruption stage (the second panel), the filament (`F1') in the western section of the AR erupted, and the subsequent disturbance propagated along the southern section of the AR to result in a typical two ribbon flare. The dark volume above the AR looks like a cavity between open and closed field lines increasing in height with time, which seemingly indicates the expansion of the dome-like separatrix. The third panel shows the second eruption that occurred in the eastern end of the AR. When the post-flare magnetic arcade extends to reach the eastern end of the AR, a loop (`F2') suddenly expands to the higher corona. This mild expansion is clearly visible at 171 \AA\, but unclear at 304 \AA, which suggests that it is of a hot loop rather than a filament. The rightmost panel shows the third eruption that broke out while the loop expansion in the eastern part was in progress. The location of the third eruption is only slightly displaced from that of the first eruption, and its evolving pattern is similar to that of the first one in that the filament  (`F3') is lifted up and a flare follows underneath.

Used as background image in the bottom panels is an HMI longitudinal magnetogram, which reveals the positive polarity fields isolated in the center of the active region and surrounded by the negative polarity fields around. This is a well known characteristic for ARs that form a dome-like separatrix structure and host circular ribbon flares (Masson et al. 2009). In the first panel we plot, as purple contours, the dark filament before eruption over the magnetogram. In the rest bottom panels, we plot only the enhanced 304 \AA\ intensities as contours on top of the magnetogram. Since 304 \AA\ images represent the chromospheric temperature, the enhanced intensity region can closely reveal the locations of the ribbons. We distinguish the flare ribbons by the inner (red contours) and the outer ribbons (blue contours) in reference to the PIL as is more or less closed. In the second panel, the typical two ribbons appear in the southern part, and another two ribbons in the northern part as well, although less clear.  The outer ribbons basically represent the open-closed boundary (OCB) for the magnetic field.  The three bottom panels show the OCB changed with time; the northern OCB shrank in some place, while the southern OCB kept expanding.The inner ribbon exhibit a more complicated evolution pattern yet in the form of a narrow lane. These two types of ribbon motions carry information on the changes of the coronal magnetic separatrix, which we attempt to derive in the next two figures.

\section{Ribbon Motions and Locations}

Figure 2 shows a time-distance ($t$--$d$) map constructed using AIA 304 \AA\ images. The slit is plotted as dashed lines in the upper panels with three cross symbols to mark  the distances 100, 200, 300 arcsec along the slit starting from its southern end. The two 304 \AA\ images are plotted from the two time intervals denoted as the vertical lines in the lower panel.  The $t-d$ map reveals a variety of phenomena including two eruptions appearing as rapidly changing features (F1 and F3) and the ribbons moving gradually (R1--R7). On the projected sky plane, F3 moves southward while F1, northward, respectively, and they appear moving in the opposite directions in the $t$--$d$ map. At the first eruption, the typical two ribbons, R1--R2, form and move away from the southern PIL as in the standard flare model. Another ribbon pair, R3--R4, formed around the northern PIL; R3 moved in the standard way, but R4 is faint and not moving much.  R2 and R4 are inner ribbons, and R1 and R3 are outer ribbons.  With time, R2 continues to move inward to merge R5.  R5 must be a continuation of R4, from which another  bright patch (denoted as R6) is extruded southward to merge with R2. As the merged ribbon R6 can no longer continue to move in the north-south direction, it instead extends in the east-west direction along a complicated path. This path must reflect the topology of the coronal quasi-separatrix above this AR. A new outer ribbon R7 came up while R3 is gone by this time. R7 must be connected to R1 over the fan dome. R1 kept moving southward, which implies that the fan dome was expanding in size.

The inner ribbon within the confined magnetic structure is a topologically special feature.  While in the ideal fan-spine structure (Parnell et al. 1996), the coronal null point is projected to a single footpoint on the surface, an inner ribbon with finite dimension would imply the presence of a quasi-separatrix layer (QSL) in the corona (Masson et al. 2009). In Figure 3 we plot the negative 304 \AA\ images to emphasize the ribbons at the two stages: 02:59 UT and 04:05 UT. The green contours represent the magnetic quantity called differential flux tube volume (DFTV), which is related to the magnetic field strength, $B$ and the path integral along the field line by $\int ds/B$. This is a measure for how rapidly field lines are locally squashed  (B\"uchner 2006).  We computed this quantity using the nonlinear force-free field (NLFFF) model by Wiegelmann (2004) from the HMI magnetogram at the nearest time, 01:58 UT and 02:46 UT, respectively, shown in the left and right panels.
In the left panel, the inner ribbon, R2, moves northward to approach another inner ribbon R4.  Both R2 and R4 do not match the high DFTV region in position, which implies that they are not directly connected to the QSL at this time.
In the right panel, R2 merges to R4 at the location of the enhanced DFTV region (green contours), after which the ribbons can no longer move north-south, and instead evolve along a more complicated path extending to east-west.
The path of this inner ribbon can be regarded as a piece of evidence for the existence of the closed, dome-shaped separatrix. It also indicates that the coronal magnetic system with the QSL sustained without much damage despite the multiple eruptions.
In the mean time, the outer ribbons, R1-R3-R7, that define the OCB significantly changed in location, implying that the overall fan-dome structure expanded in size and changed its orientation.

{\bf The bottom panels show selective field lines extrapolated from the common locations around the DFTV of two HMI vector magnetograms at 01:46:17 UT and 04:10:17 UT, respectively, viewed at a perspective angle. These model field lines do not reproduce every details of the AIA coronal images, but a few characteristics related to the magnetic field changes. First, the overall configuration is such that the field lines stemming from the enhanced DFTV region in the middle diverge to either north or south, which roughly represents the fan-dome structure. Before the eruption, the field lines connecting to southern part were more sheared (left panel) and later became somewhat relaxed (right panel), consistent with the change of R1 location. The field lines connecting to R7 reach farther to the north at 04:10 UT than at 01:46 UT. They altogether mimic the observed expansion of the OCB.}

\section{A Proposed Explanation}

We come back to the main question: how a filament could erupt without breaking the magnetic separatrix using the illustration shown in Figure 4. To facilitate a simple interpretation we start with so-called standard flare model (Priest \& Forbes 1992) plotted in the left. The magnetic field lines are plotted as solid lines, and the dashed lines are PILs.  The footpoints of the field lines form the OCB.  Two sets of two filaments are shown as thick gray lines; the smaller ones running underneath the closed field lines will have to break out the close separatrix at eruption. On the other hand, the larger ones lying above the closed field lines may freely escape to space passing through the separatrix, and the overall topology of the system structure may not change so much that similar eruption can recur as the overall structure.

In the rest two panels we apply this analogy to the fan structures. We suppose that the fan structure can be regarded as a series of the standard two-dimensional models as being wrapped up along a closed PIL so that the open field lines are merged into the spine fields and the closed field lines can form the dome-like fan structure.
A naturally arising constraint is that the PIL should lie inside the fan dome and, at least, one footpoint of a filament should lie inside the fan dome or OCB. Otherwise no filament can stand completely outside the OCB where there is only single magnetic polarity. This leaves only two cases possible as shown in the middle and right panels of Figure 4. Either two footpoints of a filament completely inside the OCB (the middle panel) or only one footpoint of the filament inside the OCB (the right panel), depending on how close the PIL lies to the OCB. In the former case a successful eruption has to break out the overlying structure, and only single eruption will be possible. In the latter case, the filament lying close to the OCB can slip through the separatrix without much altering the magnetic separatrix. Damage of the overlying separatrix to some extent by the rising filament is inevitable, but can be minimal in this case as compared with other cases where the erupting magnetic loop has a large incident angle to the separatrix. Obviously the null point can better survive in the latter scenario, which is thus favorable for the multiple filament eruptions from the confined structure.

\section{Concluding Remarks}

We have studied the  2012 March 27 eruptions from NOAA 11444  with special interest in the repeating eruptions from a confined magnetic structure. Based on the analysis of the coronal images and ribbon motions, it is argued that the multiple eruptions were possible because the topology of the  fan-spine structure was maintained during the eruptions in spite of the significant changes in its geometry.
It is also known that null-point reconnection does not necessarily blow up the fan-spine structure, and can instead produce either a remote brightening or jets (Masson et al. 2009).  Presented in this letter is, however, the type of eruptions that actually have to push through the overlying fan-spine structure or the quasi-separatrix from below on their passage. In this case it was puzzling why the perturbation at the null did not results in a drastic change in the coronal magnetic structure. In this regard, we notice a few properties by which  this event can be distinguished from other eruptions from such a fan-spine structure.

First, the magnetic polarity distribution of NOAA 11444 has a subtle difference from that of other known sources of circular ribbon flares that more often than not consist of a sigmoid AR and another neighbor AR that provides a pair polarity (Liu et al. 2015, Joshi et al. 2015). Perhaps in the latter case the sigmoid eruption produces stronger disturbance of the whole system enough to blow off the fan dome structure, resulting in a single violent eruption and strong flares, e.g., GOES M--X class flares. In contrast, the present AR has the magnetic polarity distribution more or less symmetric, but the complicated inner ribbon evolution suggest the reconnection in the quasi-separatrix in the corona, and produced two C-class flares.
Second, this AR had a distinctively long filament that erupted part by part in the outskirt of the AR (Figure 1).
Using the analogy of the standard two ribbon flare to the circular ribbon flare structure, we argued that a filament, as running nearly parallel to the separatrix, can slip through the separatrix more easily than any other magnetic structure intruding the separatrix at a larger incident angle (Figure 4). This scenario may also help explaining other cases where a magnetic null is found to be robust to disturbances (e.g., Luoni et al. 2007). Third, the dome-like separatrix rather expands with subsequent eruptions, as more field lines are reclosed after each eruption, in view of the evolution of the ribbons (Figures 2 \& 3). This is in contrast with another scenario for multiple eruptions in which magnetic reconnection proceeds from outside toward inside and the closed field lines are gradually open, like peeling off an onion (Moore et al. 2010).

To our knowledge, these eruptions represent a new type of magnetic restructuring that have not been much attended before. Instead of typical null-point reconnection or slip running reconnection often studied with circular ribbon flares, we found that sympathetic filament eruptions  occur from the confined structure and the fan-dome structure is actually reinforced after each eruption. It is argued that the successful multi-eruptions in this event is due to the interaction of the long filament encircling the AR with the fan-dome to escape from it. The role of a long filament residing under the fan-dome and the evolutionary patterns of the inner ribbons found in this study are new to the studies of circular ribbon flares, which provide an insight into the productivity of eruptions from a confined magnetic structure.

\acknowledgements
This work was supported by the National Research Foundation of Korea (NRF-2012 R1A2A1A 03670387).
J.L. was also supported by the BK21 Plus Program (21A20131111123) funded by the MOE and NRF of Korea.
C.L. is supported by NASA grant NNX13AF76G, NNX14AC12G, and NNX16AF72G, and
J.J.  by NSF grant AGS-1348513 and NASA grant NNX16AF72G.

\newpage
\begin{figure*}[tbh]  
\includegraphics[scale=0.9]{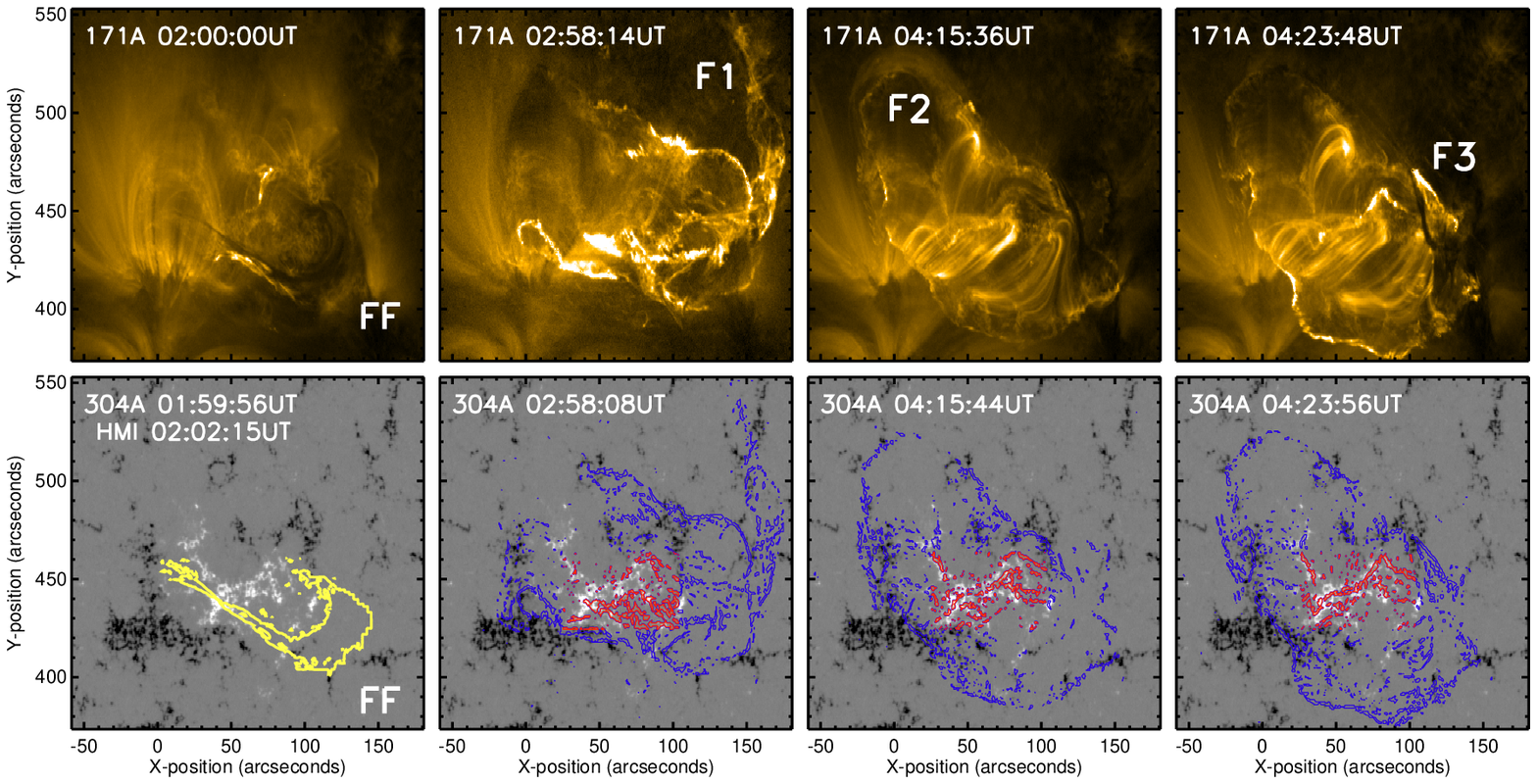}
\caption{Eruptions from NOAA 11444 on March 27, 2012.  {\it Upper panels}: AIA 171 \AA\ images at a pre-eruption stage and at three consecutive eruption stages from the left to right panels.  {\it Lower panels}: the enhanced part of AIA 304 \AA\ intensities plotted as contours on top of a pre-eruption HMI magnetogram. The upper and lower images match each other closely in time. The filaments are denoted in the upper panels and the flare ribbons, in the lower panels.}
\end{figure*}

\newpage
\begin{figure}[tbh]  
\includegraphics[scale=0.85]{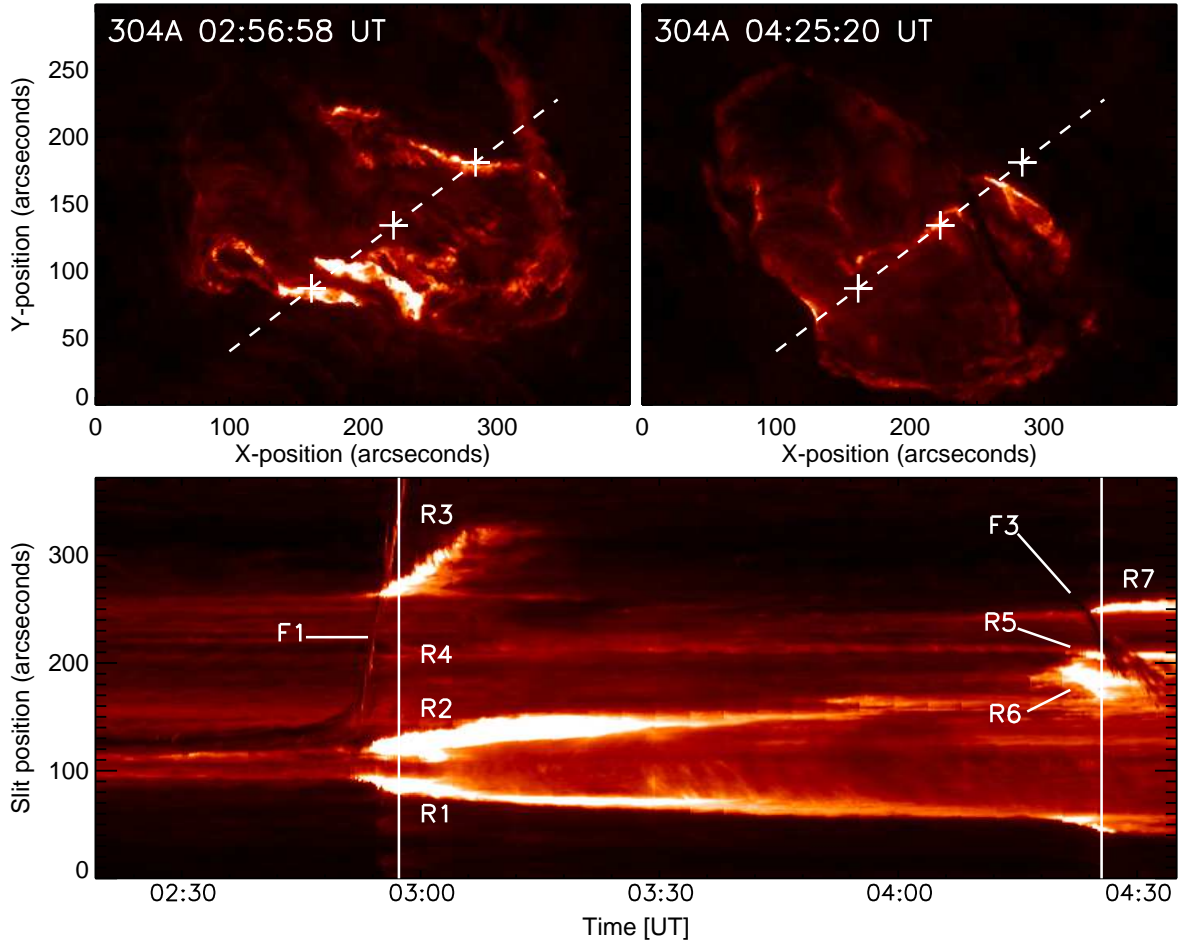}
\caption{Time-distance map (bottom panel) constructed using the slit on AIA 304 \AA\ images (top panels). {\it Top panels}: the dashed line is the slit and three cross symbols on it mark the locations of 100 arcsec interval as measured from the southern end. {\it Bottom}:  two eruptions caught on the slit are denoted as F1 and F3. R1--R7 are the ribbons. The vertical lines mark the times of the 304 \AA\ images displayed in the top panels.}
\end{figure}

\newpage
\begin{figure}[tbh]  
\includegraphics[scale=0.85]{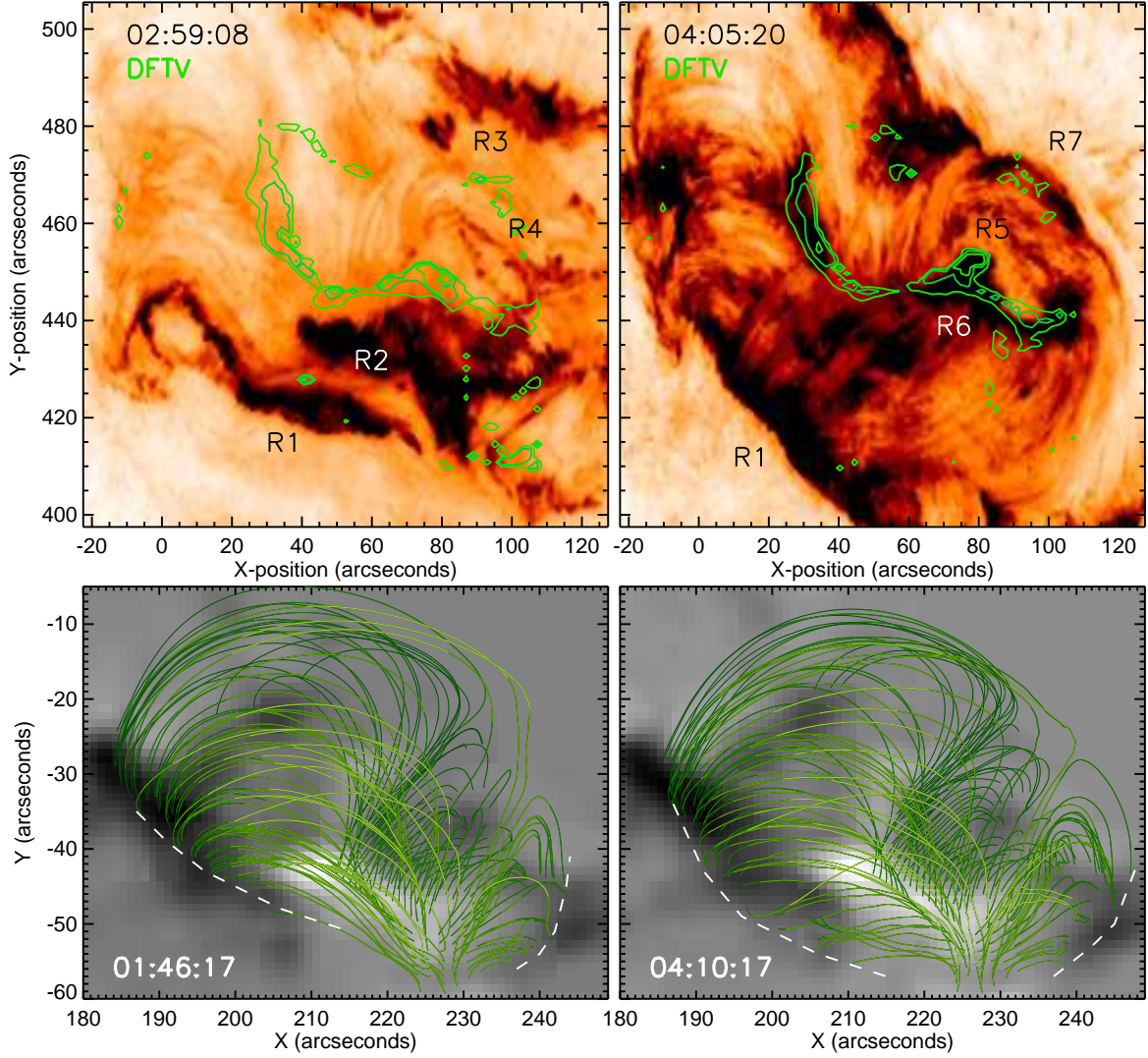}
\caption{Upper panels: locations of the inner ribbons on the negative 304 \AA\ images at the two stages: 02:59 UT (left panel) and 04:05 UT (right panel).  Green contours are the DFTV computed from the NLFFF model. Lower panels: perspective views of selective fieldlines extrapolated from the common locations around the DFTV on the HMI magnetograms at 01:46 UT and 04:10 UT, respectively. The traces of the outer footpoints of the field lines are marked with white dashed lines.}
\end{figure}

\newpage
\begin{figure}[tbh]  
\includegraphics[scale=0.85]{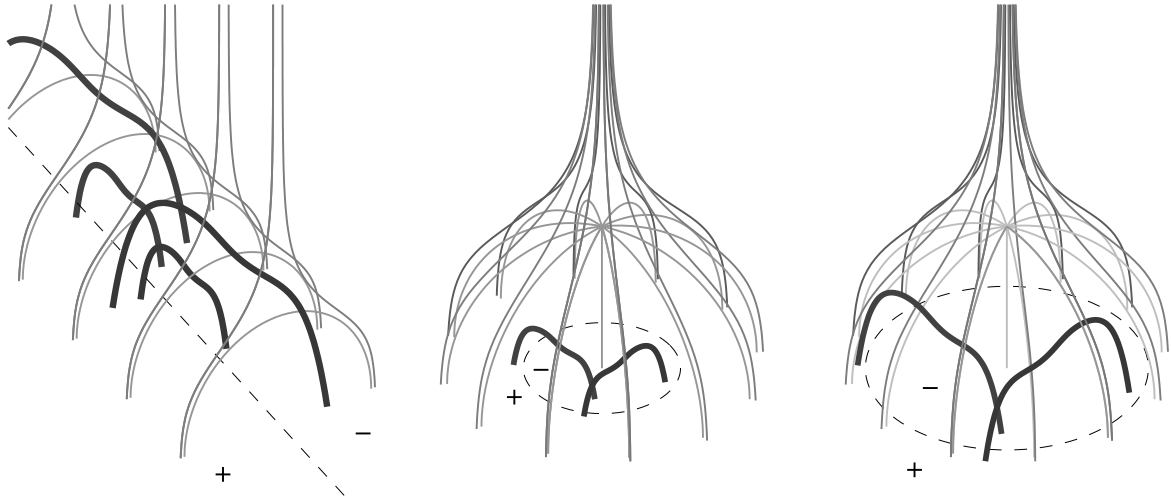}
\caption{Illustrations of the filament eruption. {\it Left panel}:  a standard two ribbon flare system with filaments above the closed magnetic loops. {\it Middle}: a fan-spine structure with filaments inside the fan-dome. {\it Right}: another fan-spine structure with filaments crossing the fan-dome. The thin lines are either the open or the closed field lines, and the thick lines represent filaments.}
\end{figure}

\end{document}